# ANALYSING THE INTERACTION OF EXPANSION DECISIONS BY END CUSTOMERS AND GRID DEVELOPMENT IN THE CONTEXT OF A MUNICIPAL ENERGY SYSTEM

*Paul Maximilian Röhrig[1,2]\*, Nancy Radermacher[1], Luis Böttcher[1], Andreas Ulbig[1,2]*

[1]*Institue for High Voltage Equipment and Grids, Digitalization and Energy Economics (IAEW), RWTH Aachen, Germany*
[2]*Digital Energy, Fraunhofer Institute for Applied Information Technology FIT, Aachen, Germany*
\* m.roehrig@iaew.rwth-aachen.de

## Abstract

In order to achieve greenhouse gas neutrality by 2045, the Climate Protection Act sets emission reduction targets for the years 2030 and 2040, as well as decreasing annual emission volumes for some sectors, including the building sector. Measures to decarbonize the building sector include energy retrofits and the expansion of renewable, decentralized power generators and low-$CO_2$ heat generators. These measures thus change both the load and the generation of the future energy supply concept. Considering the interactions of the changed installed technologies on the building level and their influence on the electrical grid infrastructure is necessary. The grid operator will remedy the future congested grid states by grid expansion measures and pass on the costs to the connected grid users, which in turn could influence their behaviour and decisions.

The aim of this work is a holistic analysis of the staggered interactions of generation expansion and grid expansion for a future decentralized energy supply concept conditioned by the expansion in the field of self-generation. To enable the analysis of the interactions, a multi-criteria optimization procedure for expansion and operation decisions at the building level is combined with an approach to determine grid expansion. As part of this work, the effect of an expansion of hosting capacity on the grid charges and thus the decision-making behaviour was investigated.

## 1 Introduction

The amended Climate Protection Act of 2021 outlines Germany's current path to achieving greenhouse gas neutrality by 2045, which includes reducing emissions by 65% by 2030 and by 88% by 2040 compared to the 1990 base year. The building sector is a major contributor to global warming, currently accounting for about 1% of all $CO_2$ emissions [1]. There is significant potential in the building sector, especially in the areas of energy efficiency and the use of renewable energy, to reduce $CO_2$ emissions and meet climate targets [1]. To effectively reduce $CO_2$ emissions, a transformation of the energy supply at the building level is subject to current planning tasks. Measures include refurbishment to improve building insulation, replacement and integration of renewable electricity and heat generators, and installation of storages [2]. This leads to an increase in decentralized energy generation close to energy consumption [3], thus leading to new electric load behaviour in the distribution grids. Furthermore, the penetration with technologies such as photovoltaics (PV) or heat pumps has high simultaneously in their load patterns which results in a change of peak load [4].

With the increase of decentralized renewable energy generation plants in the distribution grid and the decrease of conventional large power plants in the transmission grid, the previous centralized energy supply concept is changing to a decentralized approach [4]. Therefore, not only the energy transformations of individual buildings need to be considered, but also the interactions of these new plants with the electrical grid infrastructure [5]. In reality, this is a gradual process where grid operators need to reinforce lines and transformers due to the change in installed capacities [4]. The costs of these grid expansion measures are passed on to the grid users [6], which in turn can influence their decisions regarding their generation expansion.

In this paper, a method has been developed to investigate the influence of the interactions between the expansion of power generation facilities and the grid for a future decentralized energy supply concept. The highest possible spatial resolution and broad technical coverage of possible solutions are considered. The developed method uses a myopic approach to derive technology penetrations per building and considered time period and thus to map the interactions with the grid.

# 2 Methodology

In following section, we present a method to analyse the staggered interactions of end-user plant and grid expansion for decentralized energy supply concepts.

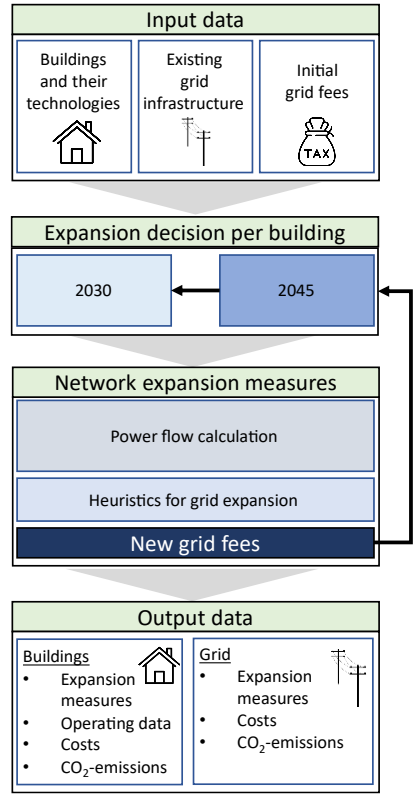

Figure 1: Overview of methodology

The model is shown in Figure 1 and is divided into four main segments: input data, building expansion decisions, determination of network expansion measures and analysis of the results.
**Input data**: Includes all relevant building-specific data, installed technologies, existing grid infrastructure and initial grid allocations.
**Building expansion decisions**: In the second block, a model for building expansion and operation decisions is applied, considering two time points: 2030 and 2045. These times were selected based on the $CO_2$ reduction targets of the Climate Protection Act [7]. Firstly, the optimum expansion measures are determined for 2045 without taking existing technologies into account. Based on this, the technologies that represent a combination of optimal technologies for 2045 and existing technologies are determined for 2030.
**Grid expansion measures**: Necessary grid expansion measures are identified for both points in time. A detailed description is provided in the grid simulation / expansion subsection. If grid expansion measures are necessary, the grid expansion costs are also calculated. The expansion planning for existing buildings is then recalculated, taking into account these additional grid expansion costs in the electricity price.
**Results analysis**: Finally, the total costs and $CO_2$ emissions of the building expansion and the grid expansion measures are evaluated. In addition, the grid expansion costs per household are calculated, based on the maximum peak load, and the expansion planning in the building stock is re-evaluated, taking these costs into account in the electricity price.

## 1.1. Building optimisation

The optimisation model (Figure 2) for the expansion and operation decisions of each building is defined as a mixed-integer optimisation problem with the objective function (1) at its core. In order to include the ecological perspective in the optimisation, we integrate a variable weighting η into the objective function, which controls the extent of the ecological consideration. A value of zero for η means a purely economic focus. [8]
The objective function is as follows:

$$\min z = (1 - \eta) * ( \sum_{i \in \{A, S\}} CAPEX_i^{fix} * x_i^{bin} + \sum_{i \in \{A, S\}} CAPEX_i^{var} * x_i^{dim} \quad (1)$$
$$+ \sum_{t \in T} \sum_{i \in \{A\}} OPEX_i * y_{i,t}^{op}) + \eta * ( \sum_{i \in \{A, S\}} NOP_i^{fix} * x_i^{bin} + \sum_{i \in \{A, S\}} NOP_i^{var} * x_i^{dim}$$
$$+ \sum_{t \in T} \sum_{i \in \{A\}} OP_i * y_{i,t}^{op})$$

$$x_i^{bin} \in \{0,1\}, x_i^{dim} \in R, x_{i,t}^{op} \in \mathbb{R}+, n \in \{0,1\} \; \forall i \in \{A, S\}, t \in T \quad (2)$$

The decision variables are defined as follows:
- $x_i^{bin}$ : Binary variable representing the investment decision in favour of building insulation refurbishment measures (S) or technologies/systems (A).
- $x_i^{dim}$: Real variable, which indicates the dimensioning of the system.
- $y_{i,t}^{op}$: Operating variable of a technology at any point in time

In order to calculate the objective function coefficients, we use the annuity method to distribute the investment and operating costs as well as the CO₂ emissions over the life cycle of the systems. The target function coefficients include fixed and variable annuities of the investment costs ($CAPEX_i^{fix}, CAPEX_i^{var}$) and the operating costs ($OPEX_i$). For the ecological dimension, fixed and variable annuity emissions ($NOP_i^{fix}, NOP_i^{var}$) and operating emissions ($OP_i$) are taken into account.

The optimisation problem includes various technical and economic constraints that can be applied to both the expansion and the operating problem. A detailed explanation of the constraints and the modelling of the problem is given in the corresponding publication [8].

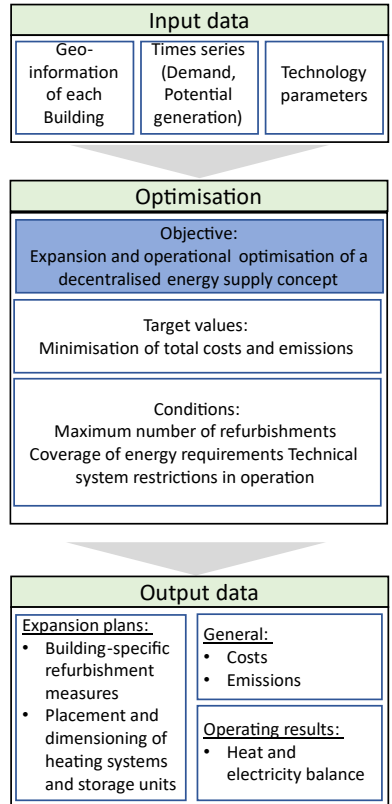

Figure 2: Overview of optimisation model for expansion and operating decisions per building

## 1.2. Grid simulation and expansion

The following section provides an overview of the power flow calculation. These results are then used to define impermissible operating states. Finally, an expansion heuristic for eliminating the bottlenecks is presented. A schematic of the method is showed in Figure 4.

**Power flow calculation**
To model the distribution grid, the power flow calculation was carried out using the open-source software MatPower [9]. The input data included information on lines and nodes as well as the specific feed-in or consumption of electrical energy at each node at each point in time. The transformer, as the interface to the medium-voltage level, was modelled as a slack node, which defines the reference voltage and can theoretically have infinite power [9]. This modelling made it possible to calculate the power flow in the grid for each point in time and to determine the energy balances at the slack node.

**Identification of unauthorised operating states**
The power flow calculation was used to create time series of all relevant state variables. Analysing this data made it possible to determine loads and their influence on the superordinate voltage level and to identify impermissible operating states, such as voltage band violations. According to the standard, the voltages at the connection points may not deviate by more than 10 % from the nominal voltage of 400 V [4]. Due to the voltage regulation at the MV level transformer, the voltage band is divided between the MV level, the local network transformer and the LV level. For this reason, a maximum permitted deviation from the voltage at the slack node of 4 % is defined in this model for each building node of the LV. [4] In addition, the thermal limits of the equipment must not be exceeded. In reality, the equipment can withstand brief exceedances of the limit values. However, due to the time resolution of one hour, the dynamics of short-term temperature changes are neglected and exceeding the nominal values in one time step is immediately categorised as an overload. [10]

**Eliminating bottlenecks through grid expansion**
When identifying impermissible operating states, heuristics were applied to ensure the restoration of permissible grid operation. In reality, the first step is to check whether operational measures can avoid grid expansion. Due to the consideration of long-term expansion decisions in the building stock and their influence on the grid, the possibility of applying these operational measures is neglected in this work.

**Grid expansion heuristics for the lines**
In the LV grid, the lines traditionally radiate from the transformer station. Due to this topology, inadmissible operating states can be directly assigned to individual lines in terms of voltage level and currents. As a network reinforcement measure, the line with the unauthorised operating state is disconnected at one point and connected to the transformer via a new parallel cable, which then supplies the connections downstream of the disconnection point. In this model, the disconnection takes place after two thirds of the overloaded line. [4] For simpler management, the same cable type and cross-section is used as for the overloaded cable. As shown in Figure 3, this measure splits the overloaded line into two lines.

In an iterative process, a parallel cable is added between the transformer station and the disconnection point and then tested using network simulation until no more voltage band violations occur. However, not all voltage band violations can be eliminated with this heuristic, which is why the maximum number of parallel cables is limited to three. In addition, this heuristic may not be able to eliminate all thermal overloads due to strong currents. This is accepted as the primary aim of this inventory model is to estimate the expansion costs.

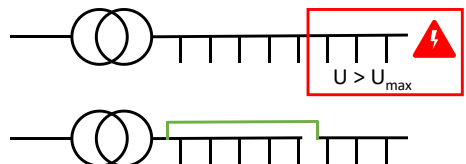

Figure 3: Expansion heuristics for the line [4]

**Reinforcement of the transformer station**
To calculate the reinforcement of the transformer station, the current flows from the grid simulation are used instead of the current load profiles for the basic consumption of the buildings. If the maximum transmitted apparent power exceeds the operating limits of the transformer, a new suitable transformer size must be selected, or two new transformers must be connected in parallel.

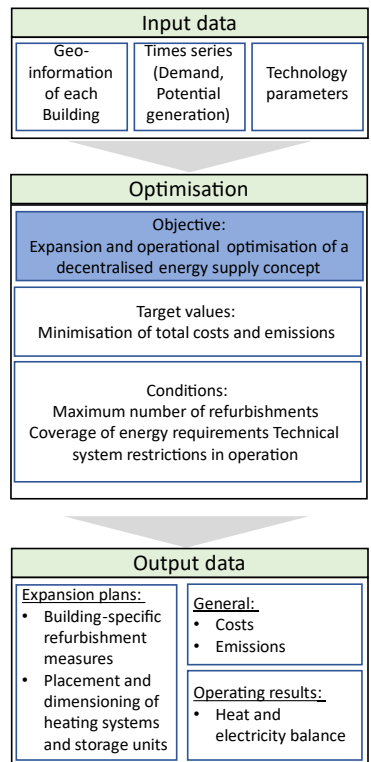

Figure 4: Overview of optimisation model for expansion and operating decisions per building

## 1.3. Distribution of grid expansion costs

Grid fees represent a significant component of energy costs that cover the apportionment of expenses for the construction, operation, and maintenance of electrical distribution grids. In Germany, the calculation of these charges is governed by incentive regulation and the Electricity Grid Charges Ordinance.

One critical aspect of incentive regulation is the use of historical costs from the base year to determine the revenue cap. This practice can lead to a delayed adjustment to current cost developments. In addition, the Electricity Grid Charges Ordinance tends to place a greater financial burden on household customers compared to large industrial consumers, which reveals a discrepancy in the distribution of costs.

**Modelling the grid expansion costs**

The model developed makes it possible to calculate the total costs for grid expansion measures. The following formula is used to determine the average grid expansion costs per kilowatt hour (kWh):

$$\emptyset \, GEC = \frac{Grid \, expansion \, costs \, [€]}{Power \, consumption \, of \, all \, buildings \, [kWh]} \quad (3)$$

The calculation takes into account both the electrical energy drawn from the grid and the electrical energy fed into the grid by the buildings.

**Differentiation of grid expansion costs**

The load and feed-in profiles of buildings vary significantly depending on the technologies implemented. This variability leads to different loads on the grid. In order to ensure a fairer distribution of grid expansion costs, the maximum peak load or peak feed-in of each building is determined in the model. Based on these values, the buildings are categorised into five evenly distributed quantiles. The allocation to a quantile determines whether a household receives a surcharge, a discount or no adjustment to the average grid expansion costs. The specific grid expansion costs per quantile are shown in Table 1.

Table 1: Grid expansion costs per quantile

| Quantile | Share of houses in the quantile | Grid expansion costs |
|---|---|---|
| Very low peak load | 20 % | GEC = 0,6 * Ø GEC |
| Low peak load | 20 % | GEC = 0,8 * Ø GEC |
| Medium peak load | 20 % | GEC = 1 * Ø GEC |
| High peak load | 20 % | GEC = 1,2 * Ø GEC |
| Very high peak load | 20 % | GEC = 1,4 * Ø GEC |

For further analyses, the grid expansion costs per household can be added to the electricity price costs. In this way, the effects of the changed electricity prices on the expansion measures in buildings can be investigated by means of a new optimisation.

## 3 Results

3.1 Subject of the study

This paper presents a model decentralised energy supply concept based on a housing estate with 80 residential units, divided into 22 apartment blocks and 58 detached houses. The selection of the years of construction of the houses reflects the distribution of years of construction of German residential buildings, based on data from the building and housing Zensus and building completions since 2000 [11]. The years of construction were assigned to the two-house types using stochastic methods. All buildings are equipped with existing heating systems, while 14% of the houses are also fitted with photovoltaic systems [12]. A detailed list of the technological parameters used as input data for the optimisation can be found in the corresponding paper [8].

Two variants are analysed:

Reference scenario (REF): The reference scenario forms the basis for the analysis and includes the calculation of the optimal expansion decisions with the aim of minimising costs for the 2030- and 2045-time horizons.

Realistic expansion (REA): Based on the reference model, this scenario adds realistic expansion rates for heat pumps - 22% by 2030 and 56% by 2045 - in order to achieve the climate targets [13]. Furthermore, the use of gas heating systems is prohibited for the final target and oil heating systems for both target horizons. The maximum refurbishment rates are limited to 53 % by 2030 and 75 % by 2045 due to the shortage of skilled labour [14].

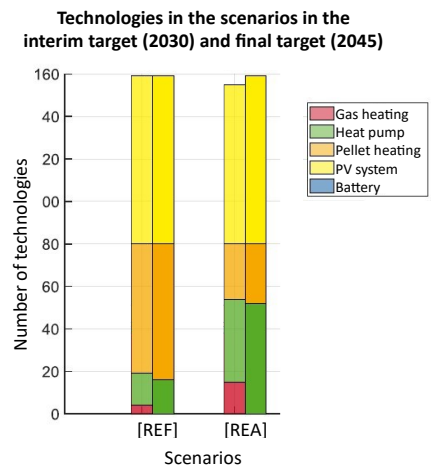

Figure 5: Expansion decision at building level

The analysis of the technology development (Figure 5) in the building sector for the year 2045 in the energy supply scenarios indicates a deployment of 64 pellet heating systems and 16 heat pumps for the reference scenario (REF). For the analysis of the effects of different expansion rates of heat pumps, this scenario serves as a starting point.

In the (REA) scenario, 39 houses are equipped with heat pumps by 2030 and a total of 52 houses by 2045, taking into account additional expansion rates of heat pumps to meet the climate targets. Compared to the (REF) scenario, this represents a significant increase of about 36 heat pumps.

Almost all houses are equipped with photovoltaic systems in both scenarios. The lack of expansion of battery storage systems in households can be explained by an economic analysis: The investment costs for batteries exceed the potential savings in operating costs that could be achieved by temporarily storing electrical energy from photovoltaic systems.

Based on these initial analyses, the influence of the levy of grid expansion costs (GEC) on expansion decisions in existing buildings will be analysed. The grid expansion costs are added to the existing electricity price as additional costs per kilowatt hour in order to assess their effect on the economic viability of expansion decisions. Two approaches are used to calculate the average GEC per kWh: one is based solely on the amount of energy drawn from the grid, while the other also takes into account the energy fed into the grid. Furthermore, it is analysed how different peak loads of the buildings influence the total costs through variable GEC surcharges or discounts.

Table 2 presents the average GEC for the (REF) and (REA) scenarios. It can be seen that the GEC in scenario (REF) is significantly lower than in scenario (REA), which is primarily due to a lower number of grid expansion measures in scenario (REF). Interestingly, the GEC are reduced by 37 % in (REF) and by 25 % in (REA) if GEC are taken into account for both energy withdrawal and feed-in to the grid.

Table 2: Average grid expansion costs per expansion path and scenario

|  | Grid expansion costs [ct/kWh] (calculation based on load and feed-in) | |
| --- | --- | --- |
| Scenario | 2030 | 2045 |
| (REF) | 0.0901 | 0.2178 |
| (REA) | 0.7480 | 0.7760 |

In previous calculations, a base value of 7.22 cents per kWh was determined. The effects of the additional GEC per expansion path on this base price are analysed below. For comparison: The average grid fees for household customers in 2021 amounted to 7.52 cents per kWh [Bun22a]. Taking only the energy load into account, the total grid fees in scenario (REF) are 1.8 % lower than the 2021 grid fees for 2030 and 0.4 % higher for 2045. In scenario (REA), the grid fees are 9 % higher than the average grid fees for 2030 and 10 % higher for 2045. However, if load and feed-in are included in the calculation, the GEC in (REF) are 2.8 % lower than the grid charges for 2030 and 1 % lower for 2045. For (REA), the grid fees for 2030 and 2045 are only 6% higher than the comparative value of 7.52 cents per kWh.

Figure 6 shows a comparison of the technology selection under the two scenarios analysed - (REF) and (REA) - taking into account the different grid expansion costs (GEC).

In the reference scenario (REF), taking the GEC into account leads to a significant adjustment in the choice of technology: The number of installed heat pumps is reduced by twelve units. This reduction illustrates how the integration of the GEC into the electricity price has a direct impact on the economic efficiency and thus on the attractiveness of heat pumps as a heating technology.

In contrast, the scenario (REA) shows that despite the increased electricity prices due to the GEC, the impact on the number of installed heat pumps is lower. With fixed expansion rates for heat pumps to achieve the climate targets, only two fewer heat pumps are installed in the final target of the (REA) scenario.

For the intermediate target of the (REA) scenario, the lower expansion rate and the permission of gas heating systems results in a more significant shift: in 24 additional houses, gas heating systems are chosen instead of heat pumps in order to minimise costs. This result shows that short-term economic considerations can have a significant influence on the choice of technology, especially when political framework conditions (such as the ban on gas heating) have not yet taken effect.

In both scenarios, taking the GEC into account in the electricity price leads to a reduced adoption of heat pumps. This observation makes it clear that GEC are a critical factor in the decision-making process for building energy technologies and can potentially impair the achievement of climate targets.

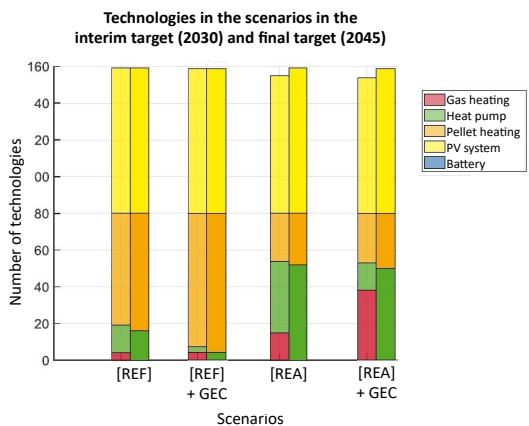

Figure 6: Expansion decision under different grid expansion costs at building level

Figure 7 shows the number of refurbishment measures per residential building under two different scenarios, namely (REF) and (REA), in relation to the varying GEC in the electricity price.

In the (REF) scenario, the inclusion of GEC in the electricity price calculation leads to a reduction in the number of refurbished residential buildings by twelve units, while in the (REA) scenario, only two fewer buildings are refurbished due to the minimum refurbishment requirement. There is a correlation between the lower refurbishment rate in the (REF) scenario and the reduced use of heat pumps, as can be seen in Figure 6. This can be explained by the fact that the heat demand in an unrenovated house cannot be covered by a heat pump. This is why at least one refurbishment measure is often carried out when a heat pump is installed. In the scenario (REF) plus changed GEC, the installation of heat pumps becomes uneconomical, which is why fewer refurbishments are required.

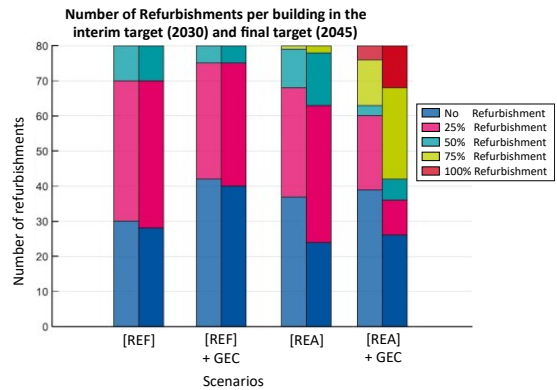

Figure 7 Refurbishment decision under different grid expansion costs at building level

The situation is different when the GEC is factored into the electricity prices in the (REA) scenario, where the number of refurbishment measures per residential building increases in both target horizons. This can be explained by the specifically defined expansion and refurbishment rates for achieving the climate targets within the (REA) scenario. Taking the GEC into account in the electricity price increases the operating costs of heat pumps, which in turn leads to an increase in refurbishment activities as a countermeasure to reduce heating costs.

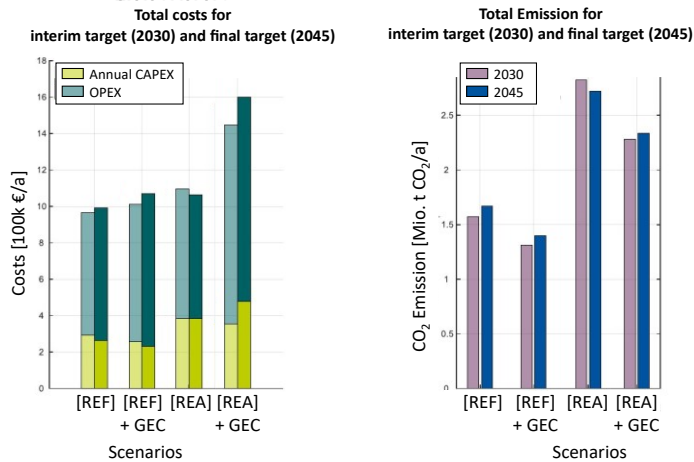

Figure 8: Total costs and total emissions

Figure 8 shows the total costs and total $CO_2$ emissions. Including GEC in the electricity price calculation results in a 13 % reduction in capital expenditure (CAPEX) for both target horizons in the (REF) scenario, which is primarily due to the reduced number of installed heat pumps. In contrast, operating costs (OPEX) in the same scenario increase by 12 % by 2030 and 16 % by 2045 due to the higher electricity price. Overall, this leads to an increase in total costs of 8 % by 2045, taking into account the GEC levy.
In the (REA) scenario, the influence of the GEC on CAPEX costs remains marginal up to the interim target, as the decrease in heat pumps is offset by an increase in refurbishment measures. For the final target, however, CAPEX costs increase by 26 % due to the constant number of heat pump installations and a significant increase in refurbishment measures compared to the baseline scenario. The total costs in this scenario increase by 53 % by 2030 and by 66 % by the final target, which can be explained by the minimum proportion of heat pumps and refurbishment measures. In terms of emissions, the inclusion of GEC in both scenarios leads to a reduction in $CO_2$ emissions of 16 % in the (REF) scenario and 19 % in the (REA) scenario by 2030 and 15 % by 2045. This is mainly due to the lower number of heat pumps used and thus the reduced electricity consumption from the grid (emission factor: 0.369 kg/kWh) [15]. In the (REA) scenario in particular, the increase in refurbishment measures leads to an additional 4% reduction in $CO_2$ emissions by 2045 due to reduced energy demand. For the year 2045, the reduced expansion in scenario (REF) results in average GEC of 0.16 ct/kWh if only the load is included in the calculation. In the (REA) scenario, the average GEC amounts to 0.35 ct/kWh for 2045, which represents an average reduction of 41 % compared to the assumed values in Table 2. However, it should be noted that convergence has not yet been achieved after two calculation runs. Theoretically, the calculations need to be continued until the average GEC added to the electricity price matches the GEC actually required.

## 4 Conclusion

In order to achieve greenhouse gas neutrality by 2045, the Climate Protection Act sets specific emission reduction targets for 2030 and 2040 and prescribes decreasing annual emission limits for selected sectors, including the building sector. The current energy supply system, which is mainly based on conventional large-scale power plants, a unidirectional energy flow and low electrification in the heating and transport sectors, needs to be fundamentally rethought. In the future system, the majority of electricity is expected to be generated from renewable sources, mainly by decentralised energy plants (DEA). The promotion of energy-efficient buildings, the switch to more environmentally friendly heating systems such as heat pumps and the increased use of photovoltaic (PV) systems have been identified as key elements for reducing $CO_2$ emissions. Due to the increasing proportion of decentralised energy producers and rising loads, expansion measures to prevent bottlenecks in the distribution grid are essential, with the costs being borne primarily by household electricity customers.

To analyse the interaction between plant expansion and grid reinforcement over several time horizons, an optimised procedure for expansion and operating decisions and a heuristic for grid expansion are introduced. These approaches take into account both optimal and existing technologies and integrate the estimated costs for grid reinforcements into decision-making processes for plant expansion.

Not taking expansion rates for heat pumps and refurbishment measures into account leads to a reduction in these technologies, whereby a scenario in which 95 % of heating systems are based on pellet heating systems and only 5 % on heat pumps by 2045 is considered problematic. This scenario shows a low level of diversification in heating technology due to the assumption of constant pellet prices and the resulting increased expansion of pellet heating systems.

By including minimum expansion rates for heat pumps and refurbishments, the expansion targets can be achieved despite an increased electricity price, but this leads to a 60% increase in operating costs. This increase should be criticised, as these costs are based on assumed expansion requirements instead of the actual expansion required. The overestimation of the grid expansion requirement due to a reduced number of heat pumps illustrates the consequences of incorrectly assumed GEC, which results in a lower expansion of heat pumps in scenario (REF) and higher operating costs in scenario (REA). Although taking GEC into account leads to a reduction in $CO_2$ emissions, it favours a one-sided heating technology and reduces the implementation of refurbishment measures that could reduce heating requirements in the long term. These dynamics require further investigation. An increase in hosting capacity with the associated increase in grid charges for a technology-open expansion decision leads to a degressive development in electrical system technologies, in order to counteract this effect it is necessary to formulate subsidies or bans on systems accordingly.

Future studies could evaluate the potential for cost reduction in grid expansion through feed-in management, flexible load control and reactive power management and analyse their influence on decisions on building expansion.